\documentclass[11pt]{article}

\usepackage{multirow}
\usepackage{epsfig}
\usepackage{array}
\usepackage{a4wide}

\begin{document}

\newcommand{\eq}[1]{eq (\ref{#1})}

\title{The scale of market quakes}

\author{
{T. Bisig$^\dag$, A.~Dupuis$^{\dag \ddagger}$, V.~Impagliazzo$^\dag$ and R.B.~Olsen$^{\dag \ddagger}$}\\
{\footnotesize ${\dag}$ Olsen Ltd., Seefeldstrasse 233, 8008 Zurich, Switzerland}\\
{\footnotesize  $\ddagger$ Centre for Computational Finance and Economic Agents (CCFEA), University of Essex, UK}
}

\maketitle

\abstract{
We define a methodology to quantify market activity on a 24 hour basis
by defining a scale, the so-called scale of market quakes (SMQ). The
SMQ is designed within a framework where we analyse the dynamics of
excess price moves from one directional change of price to the
next. We use the SMQ to quantify the FX market and evaluate the
performance of the proposed methodology at major news
announcements. The evolution of SMQ magnitudes from 2003 to 2009 is
analysed across major currency pairs.}
\newline


\renewcommand{\arraystretch}{0.2}

\section{Introduction}

Prices in financial markets evolve as events occur. Events are
typically market transactions, which may be correlated with each
other, or news and political announcements. The price evolution occurs
in many different forms and is difficult to describe concisely as price
moves happen at all different price and time scales. For instance, a
$1\%$ price move can occur within a few seconds and the price jumps to
its new level, within a few minutes and the price is subject to a
secondary counter price move or alternatively for a few days the price
may zigzag within a narrow price range. These examples highlight the
importance of creating a tool coming up with a concise abstraction
that characterises the price evolution in a systematic manner so as to
be a reliable representation of the state of the market at any point
in time and after any type of market events. We believe that a proper
tool to characterise market events is an essential element towards
building a financial warning system~\cite{olsen:09}.

The pioneering work of Zumbach~\textit{et al.}~\cite{zumbach:00} has
defined, for the currency market, the so-called scale of market shocks
that quantifies market movements on a tick-by-tick basis as a weighted
average of volatilities over different time
horizons. In~\cite{zumbach:00}, the scale of market shocks is used to
measure the impact of major events between 1998 and 1999 for a couple
of exchange rates. This approach, certainly interesting, mainly
suffers from being too complicated as too many ad-hoc choices of
functions and fittings have been made. Also we note the arbitrary
weighting of time horizons that can make the scale of market shocks to
over- or in the contrary underweight a given time-scale, therefore
distorting the measurement.

Maillet and Michel~\cite{maillet:03} have adapted the scale of market
shocks to the stock market. The new indicator is designed for the
detection and comparison of severity of different crises and suffers
from the same deficiencies as its ancestor. It is also worth
mentioning the unpublished work by Subbotin~\cite{subbotin:08} which,
also inspired by~\cite{zumbach:00}, has proposed a probabilistic
indicator for volatilities by decomposing volatility of stock market
indexes using wavelets. Wavelet decomposition is used to circumvent
the mixing of scales in~\cite{zumbach:00}. This latter indicator seems
to be usable for detecting crisis or regime shifts rather than
quantifying impact of single events as we observe periods up to a year
over which the indicator has a significant value.

For the choice of a metric to measure market evolution there cannot be
a right or wrong. There are, however, two criteria for such a metric
that seem particularly important: simplicity and ability to
incorporate all the details of the price evolution of the time
series. We claim that the previous attempts presented above have not
been able to maximise these criteria. Authors
in~\cite{zumbach:00,maillet:03,subbotin:08} have chosen volatility of
the market to characterise its state. Although this sounds like a
natural choice, we argue that aggregating market activity into a
volatility measurement is not the most appropriate as activity at
different price scales are mingled. In addition, we stress that only
the high frequency definition of volatility in~\cite{zumbach:00}
prevents information to be lost through homogenising the time series,
and even more important every time increment has the same weight.

Inspired by the discovery of a large number of scaling
laws~\cite{glattfelder:08}, we propose a framework in which price
directional changes set the rhythm and where we monitor the excess
price moves from one directional change to the next, the so-called
overshoot, at different price scales. Within this novel and simple
framework, physical time does not exist anymore and is replaced by
intrinsic time (ticking at every occurrence of a directional change of
price). The average of price overshoots does not exhibit the drawbacks
of volatility as representative price scales, and all active time
scales, are taken into account. Within this framework, we design a
methodology to quantify impact of multi-scale events along a scale,
the so-called scale of market quakes (SMQ) which defines a
tick-by-tick metric allowing us to quantify market evolution on a
continuous basis. Without loss of generality, we apply our methodology
to the FX market and publish real time magnitudes online
at~\cite{www-olsen-scale}.

Related literature in the field explores how transactions impact the
market (see~\cite{bouchaud:08} and references therein). In contrast,
we are here interested to develop tools to quantify in an objective
manner the trajectory of market prices evolution.

The document is organised as follows. We first describe the
methodology defining the SMQ and show various news announcement
snapshots to highlight the usefulness of the scale. We then analyse
the evolution of magnitudes along the SMQ over the years across major
currency pairs. Finally we conclude and discuss further work.

\section{Methodology}

\subsection{An event discretisation of the price curve}

It is custom to discretise the price curve along its temporal axis by
computing the return $r$ as the price difference within a time
interval $\Delta t$
\begin{equation}
r(t)=\log p(t)- \log p(t-\Delta t)
\label{eq:return}
\end{equation}
where $p(t)$ is the homogeneous, and therefore interpolated, sequence
of prices. Volatility and other statistical analysis are computed
from the time series (\ref{eq:return}).

Because market prices change at irregular time intervals, measurement
of market activity in terms of discrete $\Delta t$ needs to be
adaptive. To achieve that we propose an event based approach that
considers the sequence of price directional changes of magnitude
$\lambda$\cite{glattfelder:08,dacorogna:01}. Within that framework
time passes by unevenly: any occurrence of a directional change
represents a new \textit{intrinsic} time unit.

\begin{figure}
\begin{center}
\epsfig{file=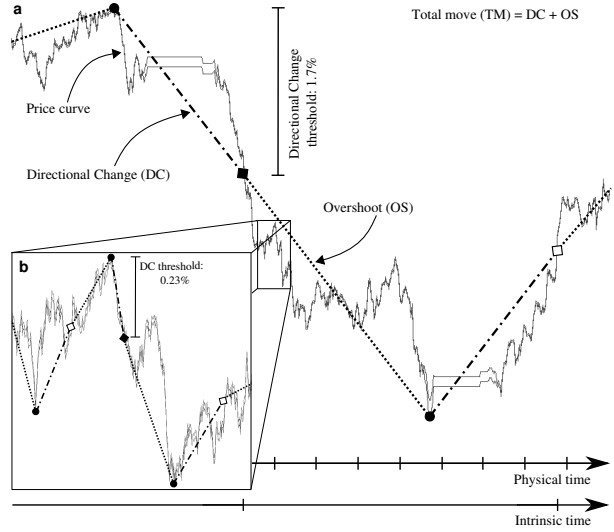,width=8cm}
\end{center}
\caption{Projection of a (a) two-week, (b) zoomed-in 36 hour price 
sample onto a reduced set of so-called directional change events
defined by a threshold (a) $\lambda = 1.7\%$, (b) $\lambda = 0.23\%$.
These directional change events (diamonds) act as natural dissection
points, decomposing a total price move between two extremal price
levels (bullets) into so-called directional change (solid lines) and
overshoot (dashed lines) sections.  Time scales depict physical time
ticking evenly across different price curve activity regimes, whereas
\textit{intrinsic time} triggers only at directional change events.}
\label{fig:os}
\end{figure}

A directional change of magnitude $\lambda$ is usually not immediately
followed by an opposite directional change but rather by a price
overshoot $\omega(t,\lambda)$ where $t$ denotes time. The price
overshoot is of particular interest as it measures the excess price
move along a $\lambda$ scale and can be used as a market activity
quantifier. Figure~\ref{fig:os} shows how the price curve is dissected
into directional change and overshoot sections.

\begin{figure}
\begin{center}
\begin{tabular}{m{1cm}m{6.5cm}}
(a) & \epsfig{file=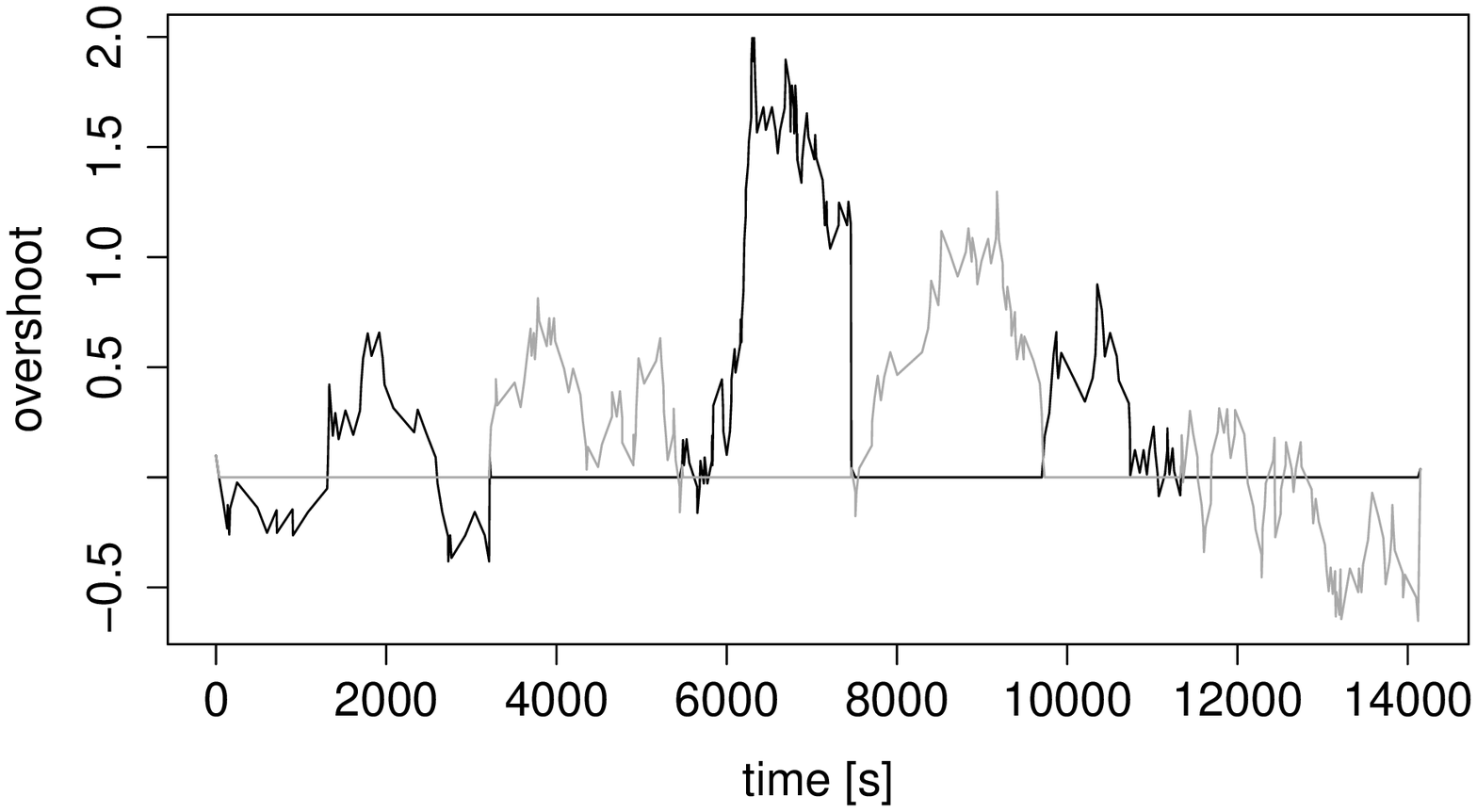,width=6.5cm} \\
(b) & \epsfig{file=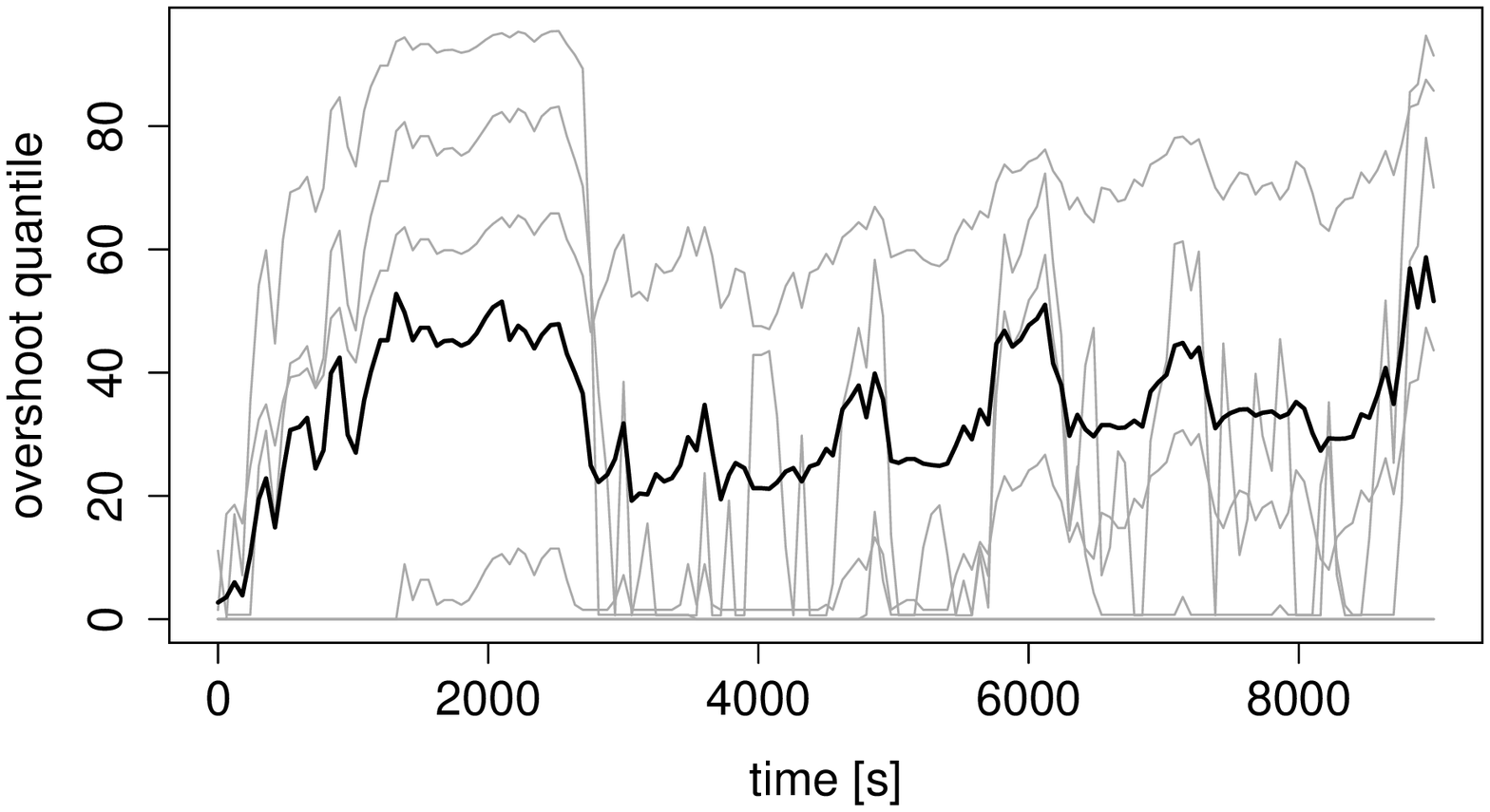,width=6.5cm} \\
\end{tabular}
\end{center}
\caption{Sample evolution of (a) the price overshoot $\omega(t)$ and (b) the
average price overshoot $\bar{\omega}(t)$. (a) Alternated gray and
black lines show the overshoot normalised by $\lambda$. (b) A subset
of the $n_\lambda$ thresholds are shown in gray and the black line
shows the average overshoot $\bar{\omega}(t)$. Overshoots are measured
in quantiles to ensure a normalised measurement.}
\label{fig:osDyn}
\end{figure}

The dynamics of overshoots organises as follows. Every occurrence of a
directional change triggers a new overshoot which will swing between
$-\lambda$ and any positive value until it decreases by $-\lambda$
making the next directional change to occur. Figure~\ref{fig:osDyn}(a)
shows the overshoot dynamics.

In order to capture the market activity, that is not occurring at a
single price scale, we define an average overshoot $\bar{\omega}$ as
\begin{equation}
\bar{\omega}(t) = \frac{1}{n_{\lambda}} \sum_{i=1}^{n_{\lambda}} \omega^q (t,\lambda_i)
\end{equation}
where $n_{\lambda}$ is the number of thresholds $\lambda_i$ and the
superscript $q$ on $\omega^q(t)$ denotes the fact that the overshoot
$\omega(t)$ is here expressed in quantiles where the tick-by-tick
historical distribution of price overshoots from December 1, 2005 up
to December 31, 2008 is considered. Overshoots are normalised by using
quantiles so as to be averaged over different thresholds. We consider
evenly distributed thresholds and set $\lambda_i = i \cdot 0.05\%$
with $i$ running from $1$ to $n_{\lambda}=100$. It is worth to note at
this stage that both $\omega(t)$ and $\bar{\omega}(t)$ are
inhomogeneous. Figure~\ref{fig:osDyn}(b) shows the time evolution of
$\bar{\omega}$.

\subsection{The scale of market quakes}

We now describe the way the average overshoot $\bar{\omega}(t)$ is
converted into a unique number $\mathcal{S}(t)$ along the scale of
market quakes. It is defined as
\begin{equation}
\mathcal{S}(t) = \frac{1}{n_a} \sum_{i=0}^{n_a} 
                 \mathcal{F}(\Omega(t+(\frac{i}{n_a}-0.5)\delta t))
\label{eq:S}
\end{equation}
where $\delta t=2$ hours is the time window, $n_a=\delta t / \delta
t_a$ with $\delta t_a = 15$ minutes and the set $\Omega(t)=\{
\bar{\omega}(\tau) - \left<\bar{\omega}(\tau)\right> \; | \; t-\delta 
t/2 \le \tau \le t+\delta t/2 \}$. The average operator
$\left<\cdot\right>$ is taken over $[t-\delta t/2 ; t+\delta t/2]$ and
is used to prevent high or low plateaux to correspond to significantly
different frequencies. The operator $\mathcal{F}(\cdot)$ is defined as
\begin{equation}
\mathcal{F}(\Omega(t)) = \frac{1}{n_f} \sum_{k=0}^{n_f-1} \frac{|X_k|}{k+1}
\end{equation}
where $n_f=\delta t / \delta t_f$ is the number of discretisation
points of $\Omega(t)$ and $|X_k|$ is the magnitude of the Fourier
frequency computed from the discretised $\Omega(t)$. $\delta t_f=7$
seconds is chosen to be precise enough but computationally doable on a
24 hour basis, and so is set to be as near as possible to a power of 2
required for an efficient computation of the Fourier transform:
$n_f=1028$ where the 4 first values are disregarded. As a discrete
Fourier transform of a signal composed of real values obeys the
symmetry $X_k=X_{n_f-k}$, we set $n_f=\lfloor n_f/2+1
\rfloor$. Dividing the frequencies by $k+1$ ensures the robustness of
the operator $\mathcal{F}(\cdot)$ to small perturbations.

The SMQ methodology is associated to a time lag of $\delta t=2$ hours
as computing a magnitude along the SMQ at time $t$ requires to know
the time series up to $t+\delta t/2$ for computing
$\mathcal{F}(\Omega(t+\delta t/2))$ and to a further $\delta t/2$ for
properly defining $\Omega$. To alleviate this issue we compute
preliminary values by reducing the averaging scope $n_a$ of
\eq{eq:S} and set $n_a = \{0,2,4,6\}$ providing estimates after
60, 75, 90, 105 minutes respectively. On average we note that
estimates are roughly $20\%$ higher than final values,
see~\cite{www-olsen-scale}.

We observe from (\ref{eq:S}) that magnitudes along the SMQ are
strictly positive. An upper bound is found by manipulating the
discrete Fourier transform definition to find a frequency bound, and
then injecting it into (\ref{eq:S}). After some simple algebra one
finds
\begin{equation}
\mathcal{S}(t) \le 50 \sum_{k=0}^{n_f-1} \frac{1}{k+1}.
\label{eq:bound}
\end{equation}
However, as we shall see, $\mathcal{S}(t)$ is usually smaller than
$10$ and \eq{eq:bound} hardly reaches its limits as it corresponds to
a theoretical case.

\section{Results}

We use the SMQ to analyse the market on a 24 hour basis and evaluate
the performance of the proposed methodology at major news
announcements in the FX market. Then we examine the evolution of SMQ
magnitudes between January 2004 up to August 2009 and across different
currency pairs. We stress here that the SMQ methodology is also
applicable to other asset classes.

\subsection{Quakes at event time}

\begin{figure*}
\begin{center}
\begin{tabular}{cc}
\epsfig{file=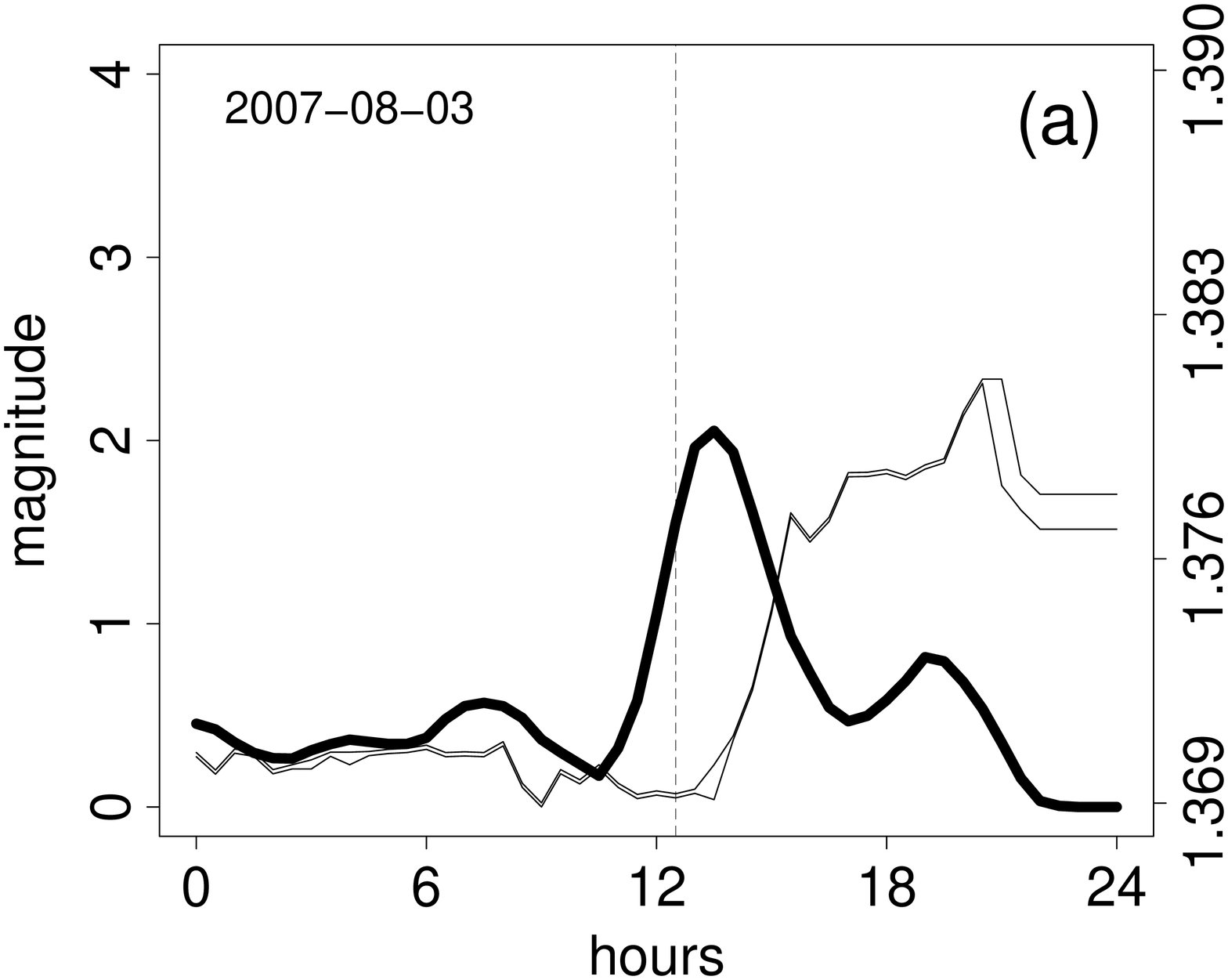,height=4.6cm,angle=-90} & 
  \epsfig{file=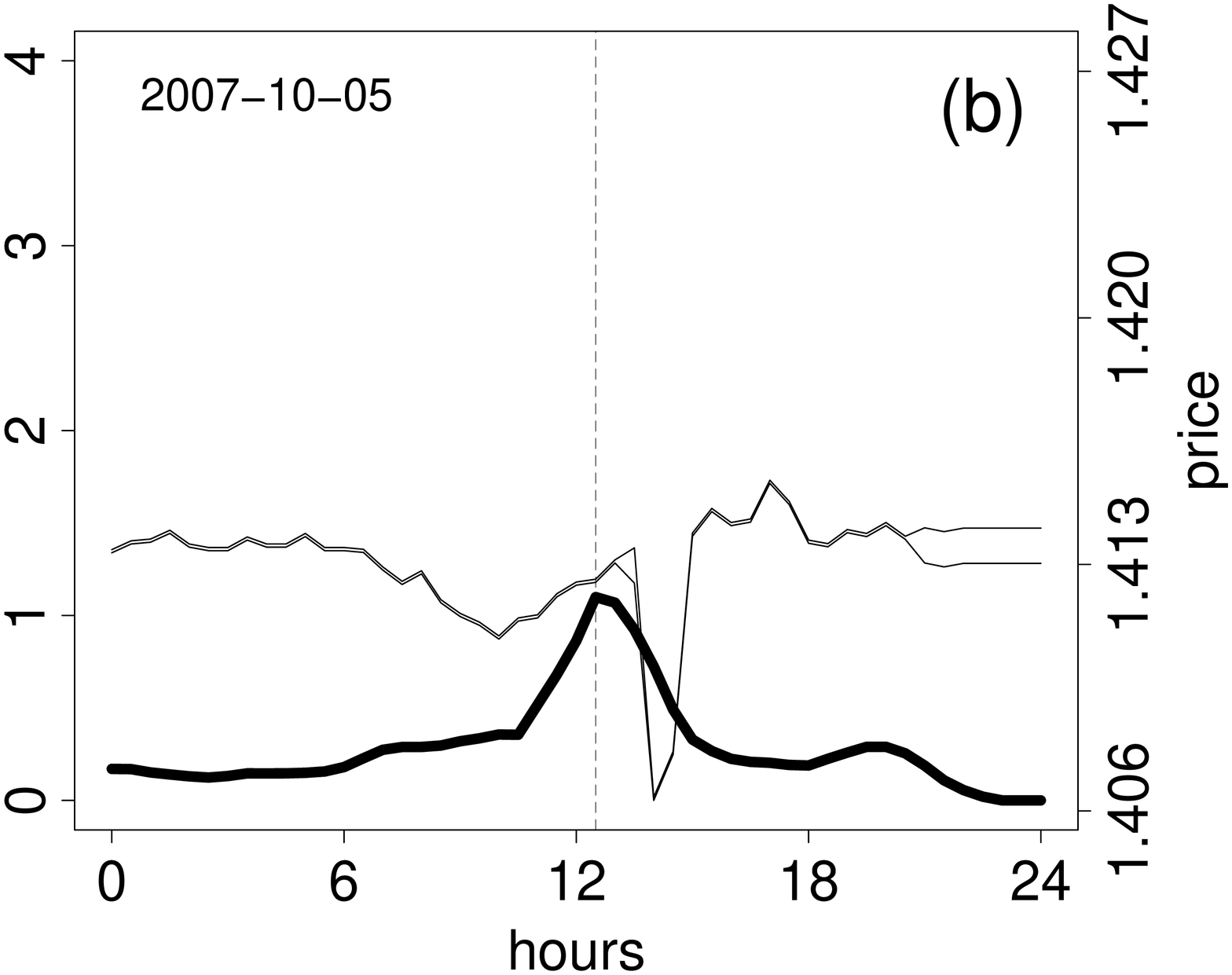,height=4.6cm,angle=-90} \\
\epsfig{file=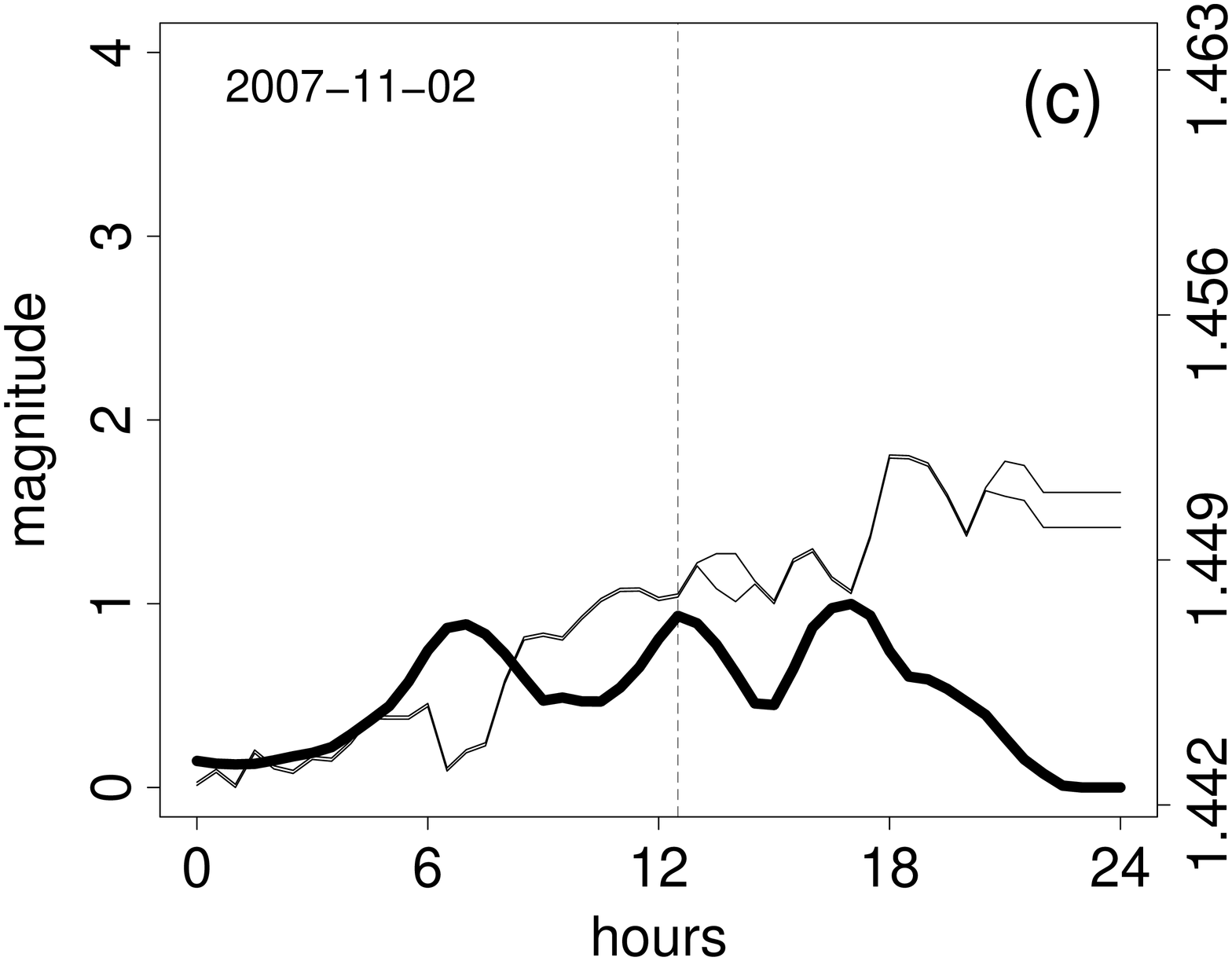,height=4.6cm,angle=-90} &
  \epsfig{file=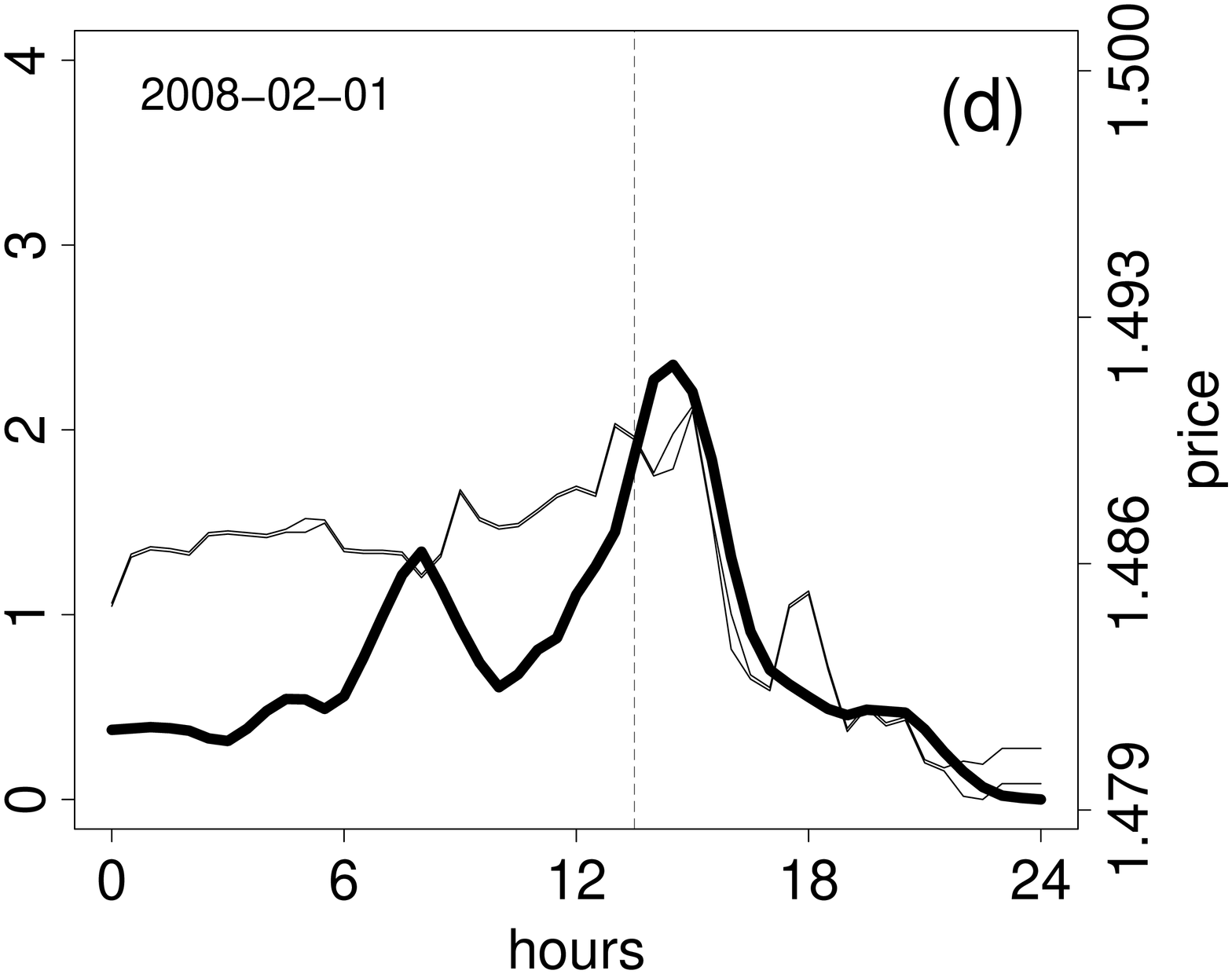,height=4.6cm,angle=-90} \\
\epsfig{file=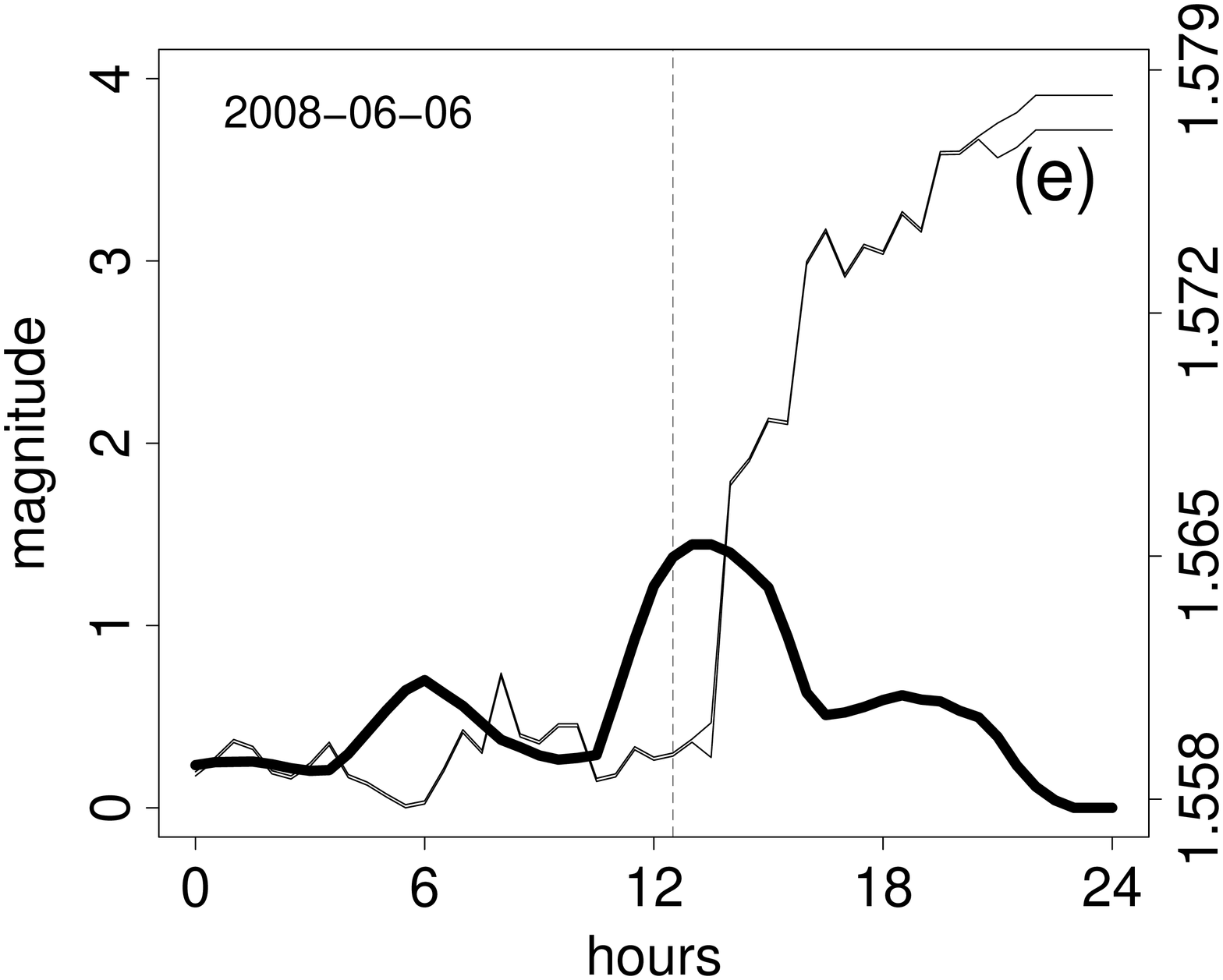,height=4.6cm,angle=-90} &
  \epsfig{file=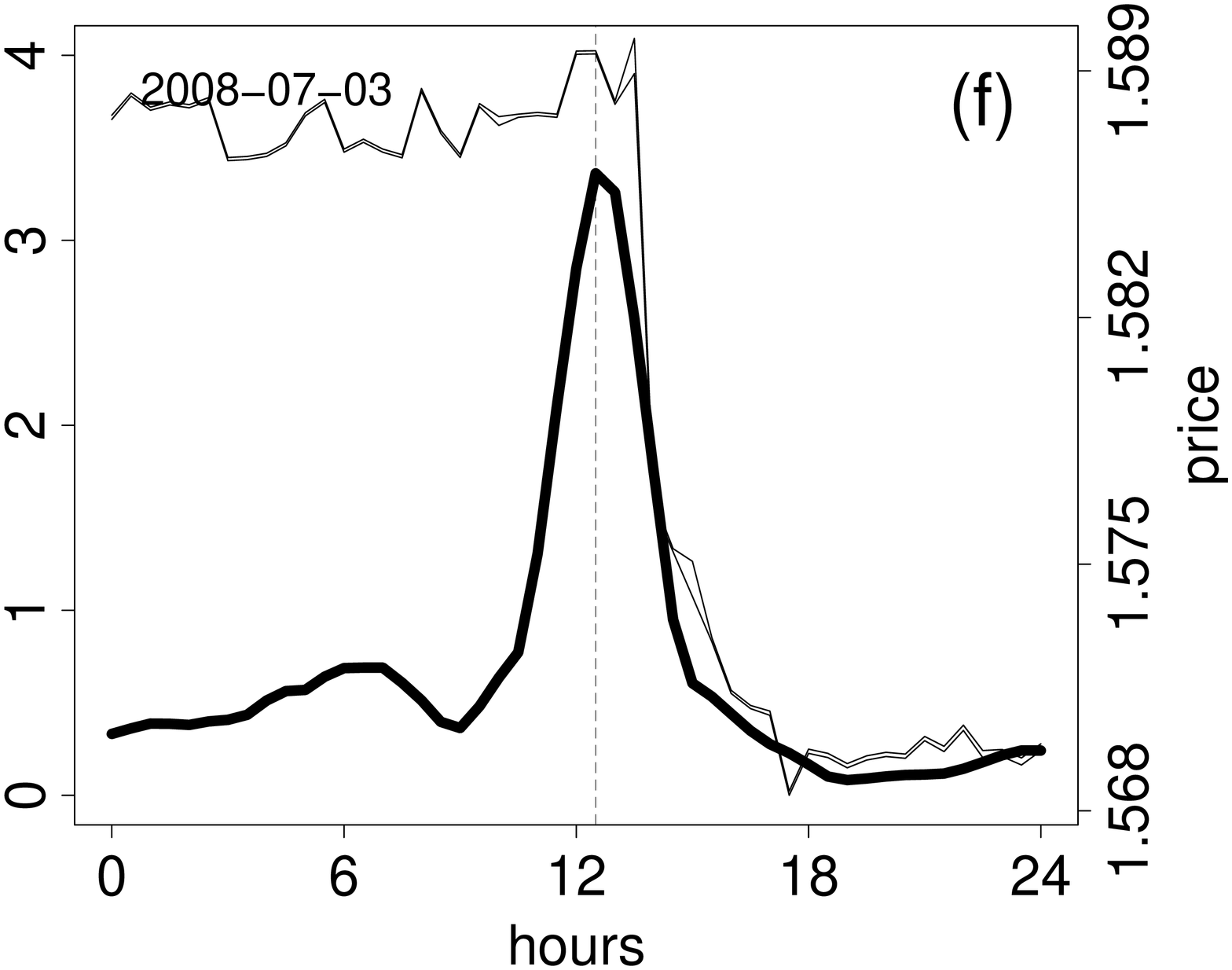,height=4.6cm,angle=-90} \\
\epsfig{file=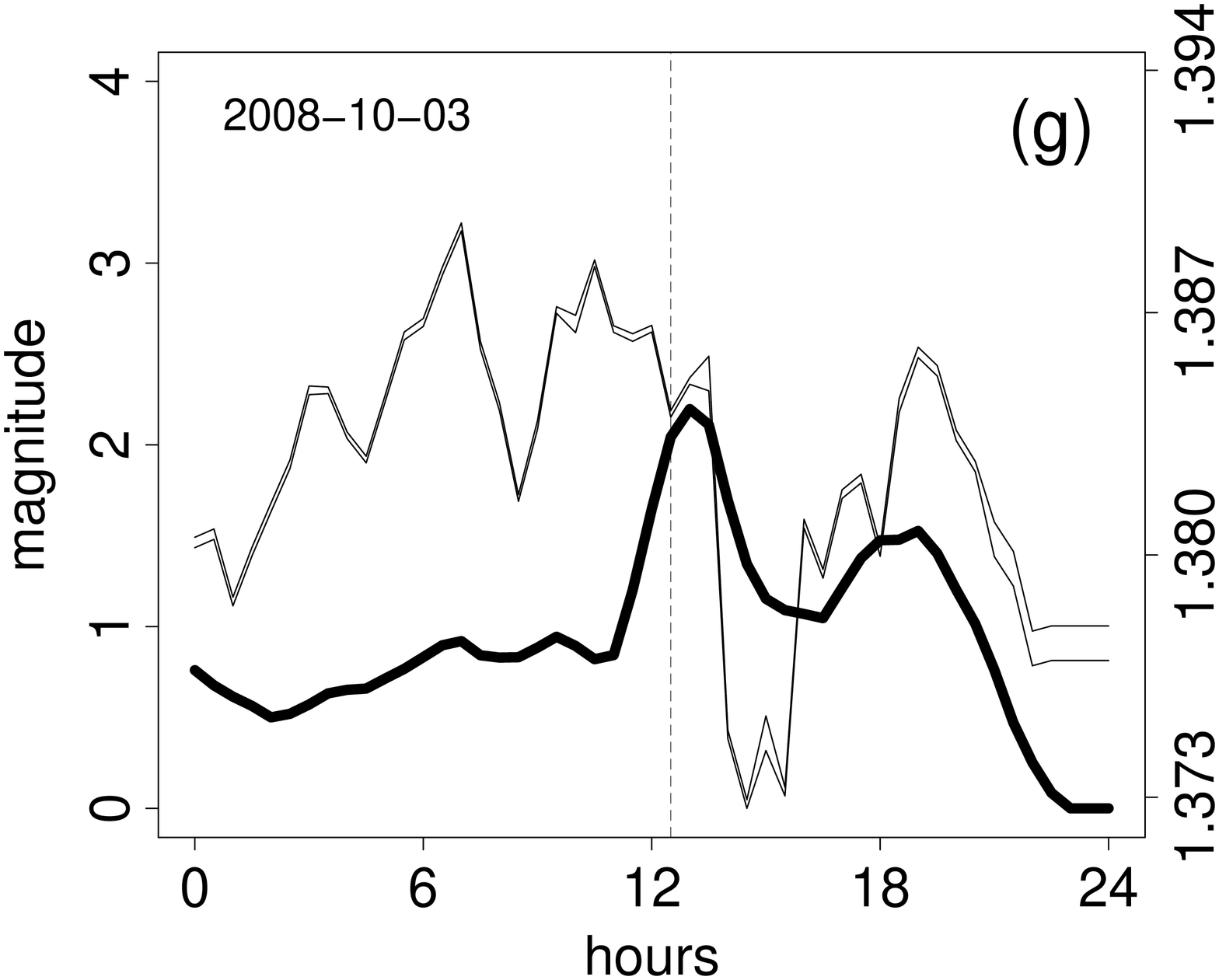,height=4.6cm,angle=-90} & 
  \epsfig{file=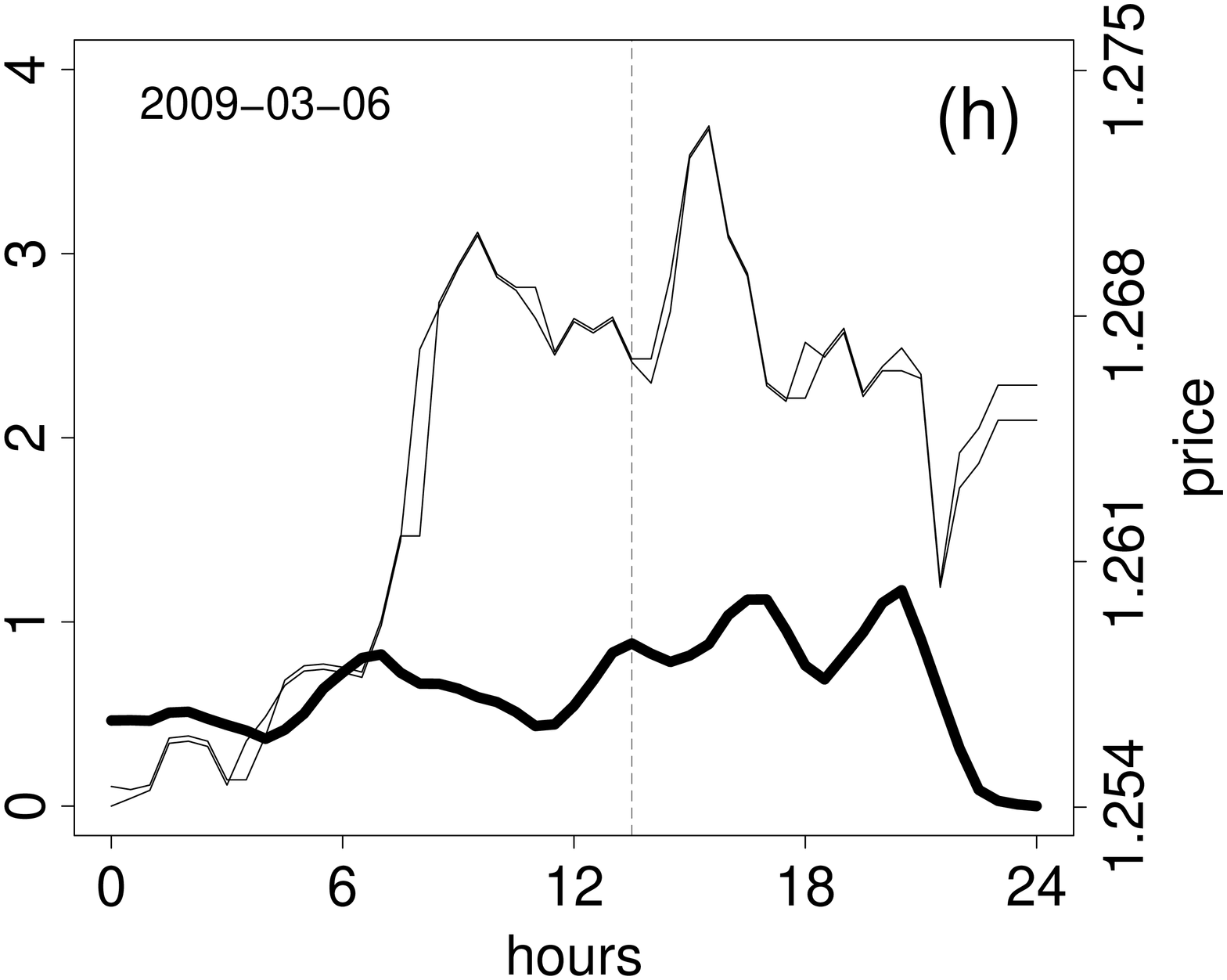,height=4.6cm,angle=-90} \\ 
\end{tabular}
\centerline{
\epsfig{file=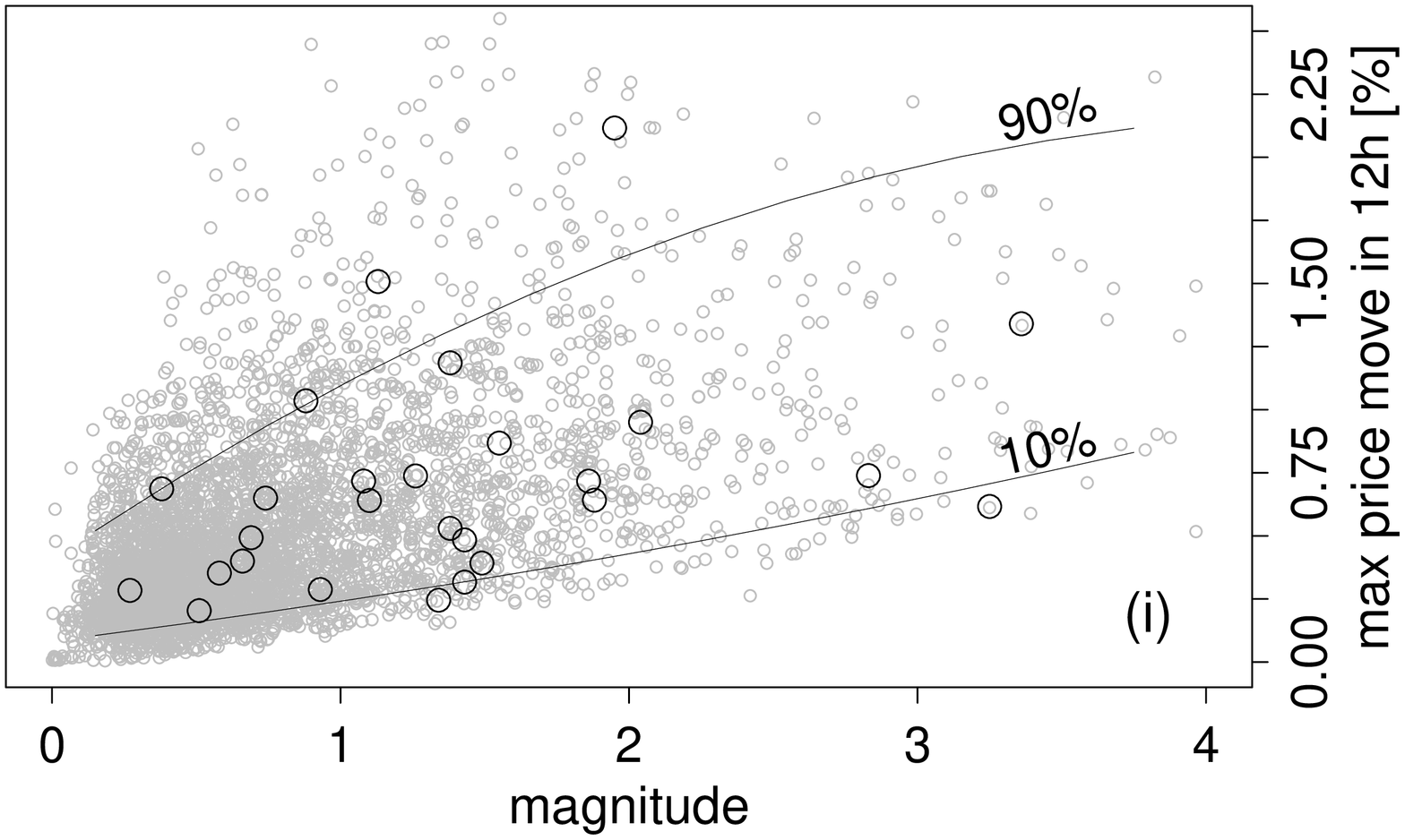,height=7.5cm,angle=-90}}
\end{center}
\caption{(a-h) Behaviour of EUR-USD (thin lines) and the SMQ 
(thick line). The announcement time is depicted by a dashed line and
its date appears on the top left of the figure. (i) Distribution of
the magnitude of two sets of events versus the maximum price move that
occurred within the next 12 hours following the events. The first
events (black circles) are 27 non-farm employment change announcements
between 2007 and 2009, and the second ones (gray circles) are 4687
magnitude peaks observed between December 2005 and March 2009 where a
magnitude peak $\mathcal{S}(t)$ is defined as $\mathcal{S}(t) >
\mathcal{S}(t\pm 2\delta t_a)$. The $10\%$ and $90\%$ quantiles of the
distribution are shown.}
\label{fig:news}
\end{figure*}

Figure~\ref{fig:news}(a-h) shows the behaviour of EUR-USD and the SMQ
on the occasion of 8 releases of non-farm employment
numbers~\cite{bls}. The wide variety of market responses: a steep drop
(f), the same price move amplitude as in (f) but happening within a
longer time period (e), little reaction from the market (c), volatile
market (g,h) or a drop immediately followed by a recovery (b,g) is
characterised by our methodology computing a single number within the
SMQ. As expected we observe that the steep drop (f) is associated to a
higher value than (e) where the difference between the two scenarios
is mainly the time for the price move to occur. Scenario (b), that
could well go unnoticed as the original price level does not seem to
be altered by the news announcement, is given a significant magnitude
that is comparable to (e).

We also notice in figure~\ref{fig:news}(a,b,d) that peaks of magnitude
do not always coincide with releasing time as the market response can
take a few hours to operate.

It is also interesting to remark that similarly to earthquakes,
after-quakes occur such as in (a,g), and have, in contrast to what is
shown here, also been observed to be stronger than the original
quake. A reason for this might be that the first market reaction could
trigger further actions producing in turn market impacts of bigger
magnitude.

Figure~\ref{fig:news}(i) shows the distribution of the magnitude of
two sets of events versus the maximum price move that occurred within
the next 12 hours following the events. The first events considered
are 27 non-farm employment change announcements between 2007 and 2009,
and the second ones are 4687 magnitude peaks observed between December
2005 and March 2009 where a magnitude peak corresponds to a value
$\mathcal{S}(t)$ where $\mathcal{S}(t) > \mathcal{S}(t\pm 2\delta
t_a)$.

We observe a cone-like structure where large magnitudes do not
correspond to any small price moves but where large price moves can be
associated to small magnitudes. This is because a large magnitude
necessarily implies that high price thresholds are activated but on
the other hand a noticeable price move can happen as a jump in the
market and therefore does not necessarily correspond to a large
magnitude.

For the sake of the presentation we have shown here only data related
to EUR-USD and only considered US news. There are no obstacles towards
considering other currency pairs and news and we are, as we write,
applying this methodology to 24 currency pairs and publishing
magnitudes related to the main international news
at~\cite{www-olsen-scale}.

\subsection{Time evolution of the magnitude}

We now analyse the time evolution of SMQ magnitudes and define the
magnitude likelihood $l_{n_d}(t,k)$ at time $t$ as the probability to
observe a magnitude to be larger or equal than a threshold magnitude
$k$, within the last $n_d$ days
\begin{equation}
l_{n_d}(t,k)=\frac{1}{n_\rho+1} \sum_{i=0}^{n_\rho} \mathcal{N}(t-i \; 2\delta t_a)
\label{eq:like}
\end{equation}
where $n_\rho=n_d/2\delta t_a$ is the number of computed magnitudes
within the backward looking time window of $n_d$ days, and where
$\mathcal{N}(t)=1$ if $\mathcal{S}(t) \ge k$ and $\mathcal{N}(t)=0$
otherwise. In the following analysis we arbitrarily set a medium-term
time window of $n_d=100$ days.

\begin{figure*}
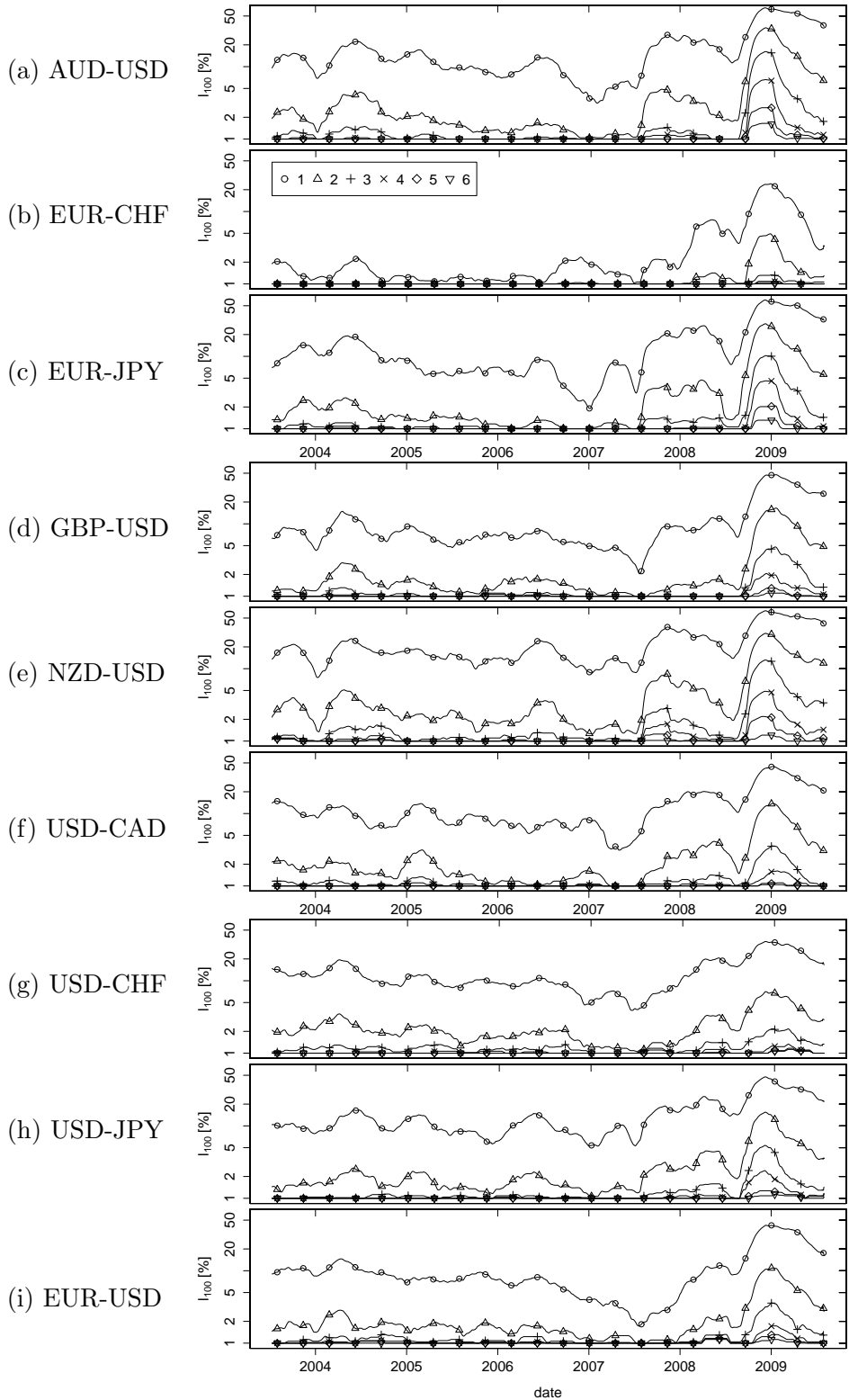

\begin{center}
\begin{tabular}{m{2.4cm}m{10cm}}
(a) AUD-USD & \rotatebox{-90}{\epsfig{file=plot_evo_AUD_USD.epsi,height=9.6cm}} \\
(b) EUR-CHF & \rotatebox{-90}{\epsfig{file=plot_evo_EUR_CHF.epsi,height=9.6cm}} \\
(c) EUR-JPY & \rotatebox{-90}{\epsfig{file=plot_evo_EUR_JPY.epsi,height=9.6cm}} \\
(d) GBP-USD & \rotatebox{-90}{\epsfig{file=plot_evo_GBP_USD.epsi,height=9.6cm}} \\
(e) NZD-USD & \rotatebox{-90}{\epsfig{file=plot_evo_NZD_USD.epsi,height=9.6cm}} \\
(f) USD-CAD & \rotatebox{-90}{\epsfig{file=plot_evo_USD_CAD.epsi,height=9.6cm}} \\
(g) USD-CHF & \rotatebox{-90}{\epsfig{file=plot_evo_USD_CHF.epsi,height=9.6cm}} \\
(h) USD-JPY & \rotatebox{-90}{\epsfig{file=plot_evo_USD_JPY.epsi,height=9.6cm}} \\
(i) EUR-USD & \rotatebox{-90}{\epsfig{file=plot_evo_EUR_USD.epsi,height=9.6cm}} \\
\end{tabular}
\end{center}
\caption{Time evolution of magnitude likelihoods $l_{100}(t,k)$ from July 2003 
to August 2009 across representative currency pairs. Nested curves
related to different magnitude threshold $k \in [1;6]$ are shown.}
\label{fig:exceed}
\end{figure*}

Figure~\ref{fig:exceed} shows the evolution of the magnitude
likelihood $l_{100}(t,k)$ for representative currency pairs, between
July 2003 to August 2009 where magnitude thresholds $k$ varies from 1
to 6. As expected from definition (\ref{eq:like}), we observe that the
smaller the threshold $k$, the larger the likelihood as
$l_{n_d}(t,k_1) \le l_{n_d}(t,k_2)$ if $k_1 \le k_2$.

Overall, figure~\ref{fig:exceed} shows decreasing likelihoods from
July 2003 up to the middle of 2007. Then the first FX market response
to the credit crisis shows: likelihoods exhibit bumps which peek at
the highest amplitude so far and vanish in the middle of 2008. These
peeks appear in the first quarter of 2008 implying that violent price
moves that occurred in summer 2007 might be responsible for these
local extrema. It then follows the second market response as we
observe sharp likelihood increases at the end of 2008. This is
followed by significant decreases over 2009 as we measure up to a
factor 4 between the maximum magnitude and the latest one in August
2009. Figure~\ref{fig:exceed} seems then to indicate that activity is
calming down in 2009, without yet indicating anything on what could
happen next. The forecasting ability of this approach is however of
interest and will be subject to further communication.

Figure~\ref{fig:exceed}(b) depicts the dynamics of EUR-CHF that seems
to behave in a somewhat different way as compared to other currency
pairs presented in figure~\ref{fig:exceed}. Indeed, magnitude
likelihoods appear to be smaller than $5 \%$ for all magnitude
thresholds, up to 2008. This is in line with a recent study by
Glattfelder~\textit{et al.}~\cite{glattfelder:08} who present a new set of 17
scaling laws that behave in a different way within EUR-CHF than within
any other of the 13 analysed currency pairs. The reason for that might
be that the ratio between volatility and spread (difference between
bid and ask) is lower in EUR-CHF than in any other currency pair
analysed here making it less attractive to speculative traders, and
therefore generating less activity.

The analysis of the evolution of likelihoods would be enhanced by
considering the average of likelihoods expressed in quantiles over all
thresholds.  A large average value would indicate an
increase of likelihood at all scales therefore highlighting the
probable start of a turbulent phase. We however leave this point for a
further study.

\section{Conclusion}

We have proposed a new and simple way of quantifying price behaviour
by computing magnitudes of quakes along the scale of market
quakes. The SMQ acts on a tick-by-tick basis~\cite{www-olsen-scale}
and quantifies multi-scale events occurring in the market in response
to news announcements or a mismatch of demand and supply. The SMQ is a
metric to measure the impact of these events.

We believe that the SMQ is a first step towards a global information
system~\cite{olsen:09} that we urgently need in order to asses the
state of the economy and its financial markets. The information system
would allow economic agents, be they governments or private and
institutional investors to take more informed decisions and adopt
preemptive action in a crisis.

\section{Acknowledgements}

We thank J.B.~Glattfelder for providing us with figure~\ref{fig:os}.

\bibliographystyle{unsrt}

\end{document}